\documentclass[sigconf]{acmart}
\usepackage{amsmath}
\usepackage{amsfonts}
\usepackage{amssymb}
\usepackage{multirow} 
\usepackage{graphicx}   
\usepackage{subcaption} 
\usepackage{multirow}
\usepackage{hyperref}

\usepackage{etex}
\usepackage{etoolbox}
\usepackage{expl3}
\usepackage{array}
\usepackage{booktabs}
\usepackage{calc}
\usepackage{euscript}
\usepackage{extarrows}
\usepackage[bottom,hang,flushmargin]{footmisc}
\usepackage{graphicx}
\usepackage{graphbox}
\usepackage{listings}
\usepackage{microtype}
\usepackage{multicol}
\usepackage{siunitx}
\usepackage{soul}
\usepackage{subcaption}
\usepackage{textcomp}
\usepackage[absolute,overlay]{textpos} 
\TPoptions{showboxes=false}
\TPMargin*{2pt} 
\usepackage{upquote}
\usepackage{verbatim}

\usepackage{catchfilebetweentags}
\makeatletter
\patchcmd{\CatchFBT@Fin@l}{\endlinechar\m@ne}{}{}{}
\makeatother

\makeatletter
\@ifclassloaded{acmart}{}{
  \usepackage{amsmath,amssymb,amsthm}
}
\@ifclassloaded{beamer}{} 
{%
  \usepackage[inline,shortlabels]{enumitem}

  \usepackage{todonotes}

  \usepackage{hyperref}
  \hypersetup{
    colorlinks=true,
    linkcolor=white,
    urlcolor=blue
    final,                   
    hidelinks,               
    breaklinks=true,         
    anchorcolor=blue,        
  }

  \newcommand{\inlinetodo}[1] {\todo[inline]{#1}\noindent}
  \newcommand{\reference}[1] {\footnotesize{\textbf{Reference:} #1}}
  \newcommand{\setlistspacing}[3] {}

  \newcommand{\showframe}[1] {%
    \begin{center}
      \framebox{\includeslide[width=8cm]{#1}}
    \end{center}
  }

  \newcommand*{\titleslide}[1] {}
}
\makeatother

\definecolor[named]{beamergreen}{rgb}{0.2,0.5,0.5}
\definecolor[named]{mygreen}{rgb}{0.4,0.9,0.7}
\definecolor[named]{myred}{rgb}{1,0.7,0.5}

\providecommand{\amber}{}
\providecommand{\blue}{}
\providecommand{\brown}{}
\providecommand{\green}{}
\providecommand{\grey}{}
\providecommand{\red}{}
\providecommand{\white}{}

\renewcommand{\amber}[1] {\textcolor[rgb]{1.00,0.50,0.00}{#1}}
\renewcommand{\blue}[1] {\textcolor{blue}{#1}}
\renewcommand{\brown}[1] {\textcolor{Mahogany}{#1}}
\renewcommand{\green}[1] {\textcolor[rgb]{0,0.7,0}{#1}}
\renewcommand{\grey}[1] {\textcolor{gray}{#1}}
\newcommand{\mygreen}[1] {\textcolor{mygreen}{#1}}
\newcommand{\myred}[1] {\textcolor{myred}{#1}}
\renewcommand{\red}[1] {\textcolor{red}{#1}}
\renewcommand{\white}[1] {\textcolor{white}{#1}}

\ifdefined\labelenumiii
  \renewcommand{\labelenumiii}{(\roman{enumiii})}
\else
  \newcommand{\labelenumiii}{(\roman{enumiii})}
\fi

\newfont{\helvetica}{phvr8t at 10pt}
\newfont{\helveticasmall}{phvr8t at 9pt}
\newfont{\helveticabig}{phvr8t at 22pt}
\newfont{\helveticao}{phvro8t at 11pt}

\newcommand{\smallsc}[1] {{\footnotesize\textsc{#1}}}
\newcommand{\setfontsizeandspacing}[2] {\fontsize{#1}{#2}\selectfont}

\makeatletter
\def\lst@visiblespace{\mbox{\kern.1em\vrule height.6ex}%
  \vbox{\hrule width1ex}%
  \hbox{\vrule height.6ex}%
}
\makeatother
\newcommand{\lstsetforpython}
{\lstset{language=Python,frame=single,frameround=tttt,framesep=3pt,backgroundcolor=\color{mygreen!20},rulecolor=\color{mygreen}}}

\usepackage{tikz} 
\usetikzlibrary{arrows.meta}
\usetikzlibrary{calc}
\usetikzlibrary{decorations.text}
\usetikzlibrary{positioning}
\usetikzlibrary{shapes, shapes.callouts}
\usetikzlibrary{trees}
\tikzset{
  >={Stealth}, 
  invisible/.style={opacity=0},
  visible on/.style={alt=#1{}{invisible}},
  alt/.code args={<#1>#2#3}{%
    \alt<#1>{\pgfkeysalso{#2}}{\pgfkeysalso{#3}} 
  },
}

\newcolumntype{C}[1]{>{\centering\arraybackslash}m{#1}}   
\newcolumntype{L}[1]{>{\raggedright\arraybackslash}m{#1}} 
\newcolumntype{R}[1]{>{\raggedleft\arraybackslash}m{#1}}  

\newcolumntype{+}{>{\global\let\currentrowstyle\relax}}
\newcolumntype{^}{>{\currentrowstyle}}
\newcommand{\rowstyle}[1]{\gdef\currentrowstyle{#1}#1\ignorespaces}

\makeatletter
\@ifundefined{htlatex}{%
  \usepackage{ctable}
  \setupctable{%
    botcap, 
    captionskip=2pt, 
    mincapwidth=2in, 
    framebg=1 1 1, 
    framefg=0 0 0, 
    framerule=0pt, 
    framesep=0pt,  
  }
}
\makeatother

\usepackage{colortbl}

\def\citea{\citet}

\newcommand{\parab}[1] {\bigskip \noindent \textbf{#1}~}
\newcommand{\param}[1] {\medskip \noindent \textbf{#1}~}
\newcommand{\paras}[1] {\smallskip \noindent \textbf{#1}~}

\newcommand{\setspacing}[1] {\linespread{#1}\selectfont}
\def\spacing {\setspacing}

\newlength{\parindentincr}
\newcommand{\addindent} {\addtolength{\parindent}{5mm}}
\newcommand{\indentby}[1] {
  \settowidth{\parindentincr}{#1}
  \addtolength{\parindent}{\parindentincr}}
\newcommand{\deindent} {\addtolength{\parindent}{-5mm}}
\newcommand{\deindentby}[1] {
  \settowidth{\parindentincr}{#1}
  \addtolength{\parindent}{-\parindentincr}}

\newcommand{\blank}[1] {\rule{#1}{0.5pt}}

\newcounter{zzlinei}
\newcommand{\blanklines}[1] {%
  \setcounter{zzlinei}{0}
  \whileboolexpr%
    {test{\ifnumcomp{\value{zzlinei}}{<}{#1}}}
    {\vspace{8mm}\hrulefill\par \addtocounter{zzlinei}{1}}
  }

\newcommand{\cl} {\centerline}
\newcommand{\command}[1] {\textcolor{Mahogany}{\texttt{#1}}}
\newcommand{\dash} {\blank}
\newcommand{\definedas} {\ensuremath{\triangleq\ }}
\newcommand{\dfslab} {\url{http://www.isical.ac.in/~dfslab}}
\newcommand{\emptybox} {\raisebox{3.3pt}{$\boxed{}$}}
\newcommand{\checkedbox} {\raisebox{3.3pt}{$\boxed{}$}\hspace{-7pt}$\checkmark$}
\newcommand{\greentick} {\includegraphics[scale=0.1]{green-tick}}
\providecommand{\half}{}
\renewcommand{\half} {\ensuremath{\frac{1}{2}}}
\renewcommand{\hl}[2][YellowOrange!50] {\colorbox{#1}{#2}}
\newcommand{\hlg}[1] {\colorbox{mygreen}{#1}}
\newcommand{\hlr}[1] {\colorbox{myred}{#1}}
\newcommand{\important}[1] {\centerline{\framebox{\textcolor{red}{#1}}}}
\newcommand{\intersection} {\cap}
\newcommand{\latex} {\LaTeX\ }
\newcommand{\lb} {\linebreak}
\renewcommand{\marks}[1] {\hspace*{\fill}{[#1]}}
\newcommand{\mc} {\multicolumn}
\newcommand{\mybox}[1] {\centerline{\framebox{#1}}}
\newcommand{\mycross} {\includegraphics[scale=0.1]{red-cross}}
\newcommand{\myellipsis}[3] {\newcount \secondindex
  \secondindex #2\relax
  \advance\secondindex +1\relax
  \ensuremath{{#1}_{#2},}
  \ensuremath{{#1}_{\the\secondindex},}
  \ensuremath{\ldots,}
  \ensuremath{{#1}_{#3}}%
}
\newcommand{\mysignature} {\includegraphics[scale=0.2]{signature}}
\newcommand{\mystrut}[1] {\rule[- #1 * \real{0.3}]{0pt}{#1}}
\newcommand{\otoprule} {\midrule[\heavyrulewidth]}
\providecommand{\overbar}{}
\renewcommand{\overbar} {\overline}
\newcommand{\partialinput}[2] {\ExecuteMetaData[#2]{#1}}
\NewCommandCopy{\pic}{\includegraphics}
\newcommand{\pipe} {~\ensuremath{\mid}~}
\newcommand{\pto} {\vfill\hspace*{\fill}\textsc{p.t.o.} \newpage}
\newcommand{\ra} {\ensuremath{\rightarrow}}
\newcommand{\Ra} {\ensuremath{\Rightarrow}}
\newcommand{\rcomment}[1] {\bigskip\color{blue}#1\color{black}}
\newcommand{\response} {\medskip\textbf{Response:~}}
\newcommand*{\rectgolla}[2][red]{
  \setlength{\fboxrule}{1pt}
  \fcolorbox{#1}{white}{#2}
} 
\newcommand{\sbcb} {\;\!} 
\renewcommand{\show} {\visible}
\renewcommand{\so} {\st}
\newcommand{\strikethrough} {\st}
\newcommand{\tb} {\textbackslash}
\newcommand{\torf} {\hfill \textsc{true / false}}
\newcommand{\tpgridon}{\TPShowGrid*{15}{12}} 
\newcommand{\ttilde} {\textasciitilde}
\ifdefined\ul
  \renewcommand{\ul}[1] {\underline{#1}}
\else
  \newcommand{\ul}[1] {\underline{#1}}
\fi
\newcommand{\undertilde}[1] {\mbox{\Large \raisebox{-5mm}{$\displaystyle\stackrel{#1}{\text{\textasciitilde}}$}}}
\newcommand{\union} {\cup}
\NewCommandCopy{\oldvec}{\vec}
\renewcommand{\vec}[1] {\oldvec{\vphantom{b}#1}}
\newcommand{\xor} {\ensuremath{\oplus}}

\newcommand{\anchorpoint}[1]{\tikz[overlay,remember picture] \node (#1) {};}

\newcommand*{\floatingnote}[5][1-]{
  \begin{tikzpicture}[overlay,remember picture]
    \pgftransformshift{\pgfpointanchor{current page}{center}}
    \path<#1> #2
    node[rectangle,rounded corners=4pt,fill=#4,opacity=.5]
    {\parbox{#3}{\raggedright #5}}
    ;
  \end{tikzpicture}
}

\newcommand*{\golla}[4][ellipse]{
  \begin{tikzpicture}[overlay,thick]
    \node () at #2 [shape=#1,draw=red,text height=#4] {~\hspace{#3}~} ;
  \end{tikzpicture}
}
 
\newcommand*{\namedgolla}[5][red]{
  \begin{tikzpicture}[overlay,remember picture,thick]
    \node (#5) at #2 [shape=ellipse,draw=#1,text height=#4] {~\hspace{#3}~} ;
  \end{tikzpicture}
}

\newcommand*{\popupnote}[4][1-]{
  \begin{tikzpicture}[overlay,remember picture]
    \pgftransformshift{\pgfpointanchor{current page}{center}}
    \path<#1> (#2) ++#3
    node[anchor=west,rectangle callout,rounded corners=4pt,fill=blue!50,
         opacity=.5,callout absolute pointer={(#2)}] {#4};
  \end{tikzpicture}
}

\makeatletter
\newread\pin@file
\newcounter{pinlineno}
\newcommand\pin@accu{}
\newcommand\pin@ext{pintmp}
\newcommand*\dummypartialinput [3] {%
  \IfFileExists{#3}{%
    \openin\pin@file #3
    \setcounter{pinlineno}{1}
    \@whilenum\value{pinlineno}<#1 \do{%
      \read\pin@file to\pin@line
      \stepcounter{pinlineno}%
    }
    \addtocounter{pinlineno}{-1}
    \let\pin@accu\empty
    \begingroup
    \endlinechar\newlinechar
    \@whilenum\value{pinlineno}<#2 \do{%
      \readline\pin@file to\pin@line
      \edef\pin@accu{\pin@accu\pin@line}%
      \stepcounter{pinlineno}%
    }
    \closein\pin@file
    \expandafter\endgroup
    \scantokens\expandafter{\pin@accu}%
  }{%
    \errmessage{File `#3' doesn't exist!}%
  }%
}
\makeatother

\AtBeginDocument{%
  \providecommand\BibTeX{{%
    Bib\TeX}}}

\setcopyright{acmlicensed}
\copyrightyear{2018}
\acmYear{2018}
\acmDOI{XXXXXXX.XXXXXXX}
\acmConference[Conference acronym 'XX]{Make sure to enter the correct
  conference title from your rights confirmation email}{June 03--05,
  2018}{Woodstock, NY}
\acmISBN{978-1-4503-XXXX-X/2018/06}

\begin{document}

\title{LLMs as Assessors: Right for the Right Reason?}


\author{Sourav Saha}
\affiliation{%
	\institution{Indian Statistical Institute}
	\city{Kolkata}
	\country{India}
}
\email{sourav.saha\_r@isical.ac.in}

\author{Mandar Mitra}
\affiliation{%
	\institution{Indian Statistical Institute}
	\city{Kolkata}
	\country{India}
}
\email{mandar@isical.ac.in}

\author{Aditya Dutta}
\affiliation{%
	\institution{Indian Statistical Institute}
	\city{Kolkata}
	\country{India}
}
\email{adi22dutta@gmail.com}



\begin{abstract}
  A good deal of recent research has focused on how Large Language Models
  (LLMs) may be used as `judges' in place of humans to evaluate the quality
  of the output produced by various text / image processing systems. Within
  this broader context, a number of studies have investigated the specific
  question of how effectively LLMs can be used as relevance assessors for
  the standard ad hoc task in Information Retrieval (IR). We extend these
  studies by looking at additional questions. Most importantly, we use a
  Wikipedia based test collection created by the INEX initiative, and
  prompt LLMs to not only judge whether documents are relevant /
  non-relevant, but to highlight relevant passages in documents that it
  regards as useful. The human relevance assessors involved in creating
  this collection were given analogous instructions, i.e., they were asked
  to highlight all passages within a document that respond to the
  information need expressed in a query. This enables us to evaluate the
  quality of LLMs as judges not only at the document level, but to also
  quantify how often these `judges' are right \emph{for the right reasons}.
  Our observations lead us to reiterate the cautionary note sounded in some
  earlier studies when it comes to using LLMs as assessors for creating IR
  datasets: while LLMs are unquestionably promising, and may be used
  judiciously to subtantially reduce the amount of human involvement
  required to generate high-quality benchmark datasets, they cannot replace
  humans as assessors.
\end{abstract}




\maketitle

\section{Introduction}
\label{sec:introduction}
A good deal of recent research has focused on whether and how LLMs may
replace human assessors or
annotators~\cite{faggioli-ictir-2023,llm_judges_bhaskar_da,lara,umbrela-ictir-25,alaofi-tois-2025}
when creating `ground truth' or `gold standard' data that is commonly used
to evaluate information access systems. \citet{faggioli-ictir-2023}
provided an early perspective on the issue, by reporting the results of
some `pilot' experiments, and describing the advantages of using
LLMs-as-Judges, while also warning against potential pitfalls. Some
subsequent studies have painted an encouraging picture of the use of LLMs
as assessors~\cite{llm_judges_bhaskar_da,umbrela-ictir-25}, others have
sounded a cautionary note~\cite{clark-dietz-arxiv-2025,Soboroff_2025},
while yet others~\cite{lara} have discussed how the best balance may be
achieved by cooperating LLMs and human assessors.

This study contributes to the ongoing debate on LLMs-as-assessors. To the
best of our knowledge, our study is the first to look beyond document-level
relevance labels in this context: we ask LLMs to \textbf{extract (or
  highlight) the parts of a document that it thinks are relevant} for the
information need expressed in a given query. These highlighted passages may
be regarded as \emph{rationales} (a term that is commonly used in the
literature on explainability~\cite[esp.\ Section
9]{explain-ir-avishek-anand-survey}) that help us to understand \emph{why}
an LLM regards a document as relevant to a query.

We focus on a variant of the canonical \emph{ad hoc search}
task~(\cite{umbrela-ictir-25}, Section~3.1; \cite{Soboroff_2025}, Section
1) in Information Retrieval (IR). Instead of retrieving complete documents
in response to queries, an IR system is required to extract and retrieve only the
relevant portions of useful documents. For our experiments, we leverage the
INEX 2009 and 2010 collections~\cite{inex_2009,inex_2010}, which provide
human-annotated ground-truth data on a Wikipedia dump for this task. We use
a `few-shot' setting by providing (via the prompt) a handful of examples
illustrating what we want. Our experiments enable us to quantify how well
LLMs agree with humans about which parts of a relevant document are
actually useful, and thereby, to (partially) address the issue of whether
LLMs are ``right for the right
reasons''~\cite{explain-ir-avishek-anand-survey,anand-tutorial-sigir-2023}.

Our observations lead us to reiterate the cautionary note sounded in some
earlier studies when it comes to using LLMs as assessors for creating IR
datasets: while LLMs are unquestionably promising, and may be used
judiciously to subtantially reduce the amount of human involvement required
to generate high-quality benchmark datasets, they cannot replace humans
as assessors. 
We find that some LLMs are overly inclined to regard documents as relevant,
while larger, more modern models are excessively conservative. Likewise,
there is a sizeable lack of alignment between human and LLM behaviour when
LLMs are asked to mark the useful portions of relevant documents. In
particular, LLMs seem to struggle with finding `needles in haystacks',
i.e., when only a small passage within a sizeable document is relevant for
a certain query. This is also true for documents that contain multiple
discontiguous chunks of useful information.


In the next section, we discuss the relation between recent studies and our
work in greater detail. Section~\ref{sec:approach} describes the main
components of our experiments, including how the INEX task and data were
used in our experiments, along with issues related to prompt structuring
and exemplar selection. Section~\ref{sec:setup} provides more details about
the experimental setup, and the LLMs used. Results are discussed in
Section~\ref{sec:results}. Finally, in Section~\ref{sec:cfw},
we 
list a number of additional questions that we intend to explore in the near
future. For reproducibility purposes, we release our codebase
at~\url{https://anonymous.4open.science/r/llm-as-judge}.


\section{Related work}
\label{sec:related-work}

\citet{chiang-lee-2023-large} claim to be ``the first to show the potential
of using LLMs to assess the quality of texts and discuss the limitations
and ethical considerations of LLM evaluation.'' They compare humans and
LLMs as evaluators for two generative tasks in Natural Language Processing
(NLP): open-ended story generation and adversarial attacks. Within the IR
community, early work on LLMs-as-assessors is represented by the
discussions in~\citet{bauer-dagstuhl-2023} and~\citet{faggioli-ictir-2023}.
In a number of
studies~\cite{llm_judges_bhaskar_da,umbrela-ictir-25,clark-dietz-arxiv-2025,
  sakai-cikm-25-short,alaofi-tois-2025}, researchers have used LLMs to
create both binary and graded relevance judgments for query-document (Q-D) pairs.
These judgments are evaluated both directly, in terms of agreement /
overlap with human-assigned labels, and indirectly, in terms of whether IR
systems are ranked differently when evaluated using judgments provided by
LLMs or humans. 

One recent study~\cite{llm_judges_bhaskar_da} presented encouraging results
by showing that LLM-based relevance assessors correlate strongly with
trained human assessors, and generally do better than crowdsourcing. On the
whole, system rankings do not change dramatically when human judgments are
replaced by automatic judgments, with high values of Kendall's $\tau$ (0.8
and higher) being reported~\cite{umbrela-ictir-25}. They do not study few-shot prompting; further, they argue that TREC documents are too long to be included in zero-shot or few-shot prompts. In contrast, our study includes Wikipedia documents whose lengths are comparable to those of TREC documents.

However, several researchers~\cite{faggioli-ictir-2023,Soboroff_2025} have
raised serious concerns --- both philosophical and empirical --- about
LLMs-as-assessors. For example, in the experiments reported in
\cite{clark-dietz-arxiv-2025}, Kendall's $\tau$ values drop to around 0.51
when the ranking of only the 20 best systems is considered; thus, automatic
judgments may not be reliable when comparing competing,
near-state-of-the-art systems.
Similarly, \citet{alofi} showed that an LLM can be misled into predicting a
document as ``perfectly relevant,'' even when it is not actually relevant
to the given query. Therefore, relying heavily on automated relevance
assessment frameworks remains risky. In a similar vein, \citet{sakai-cikm-25-short} argued that even when evaluated on reliable test collections, LLMs cannot fully replace human assessors due to limited discriminative ability; they often fail to distinguish relevance in the nuanced manner achieved by human judgment. The objective of that line of work is to examine whether LLMs can faithfully mimic human relevance assessments. They further suggest that open-source LLMs cannot be applied at the run level.

Another valid concern involves biases that may be inherent in LLMs used as
judges. \citet{JudgeBlender} propose to reduce biases from closed source
models by using an aggregation framework to combine relevance judgments
from different LLMs into a single one, but this issue needs more careful
investigation.

While fully automatic judgments represent one end of the human-LLM
collaboration spectrum~\cite{faggioli-ictir-2023}, researchers have also
considered a more restricted setting, by using LLMs only on
documents that have not explicitly been judged by humans (and which are
therefore assumed to be
non-relevant)~\cite{sean_chora,abbasiantaeb2024uselargelanguagemodels}. The
question of how to achieve a good balance between manual and LLM
annotations has been recently studied by \citet{lara}. They present a
method for calibrating LLM scores using a regression model trained on a few
graded relevance assessments. Based on this calibration, we may estimate
how confident an LLM is about a prediction. The eventual objective is to
eliminate the need for human assessors for decisions that the LLM is
confident about, while using human annotations in cases where the LLM
exhibits low confidence.

All the work mentioned thus far considers relevance at the level of full
documents only. Even though some of the above studies use ``passage''
collections, e.g., MS MARCO 
for their experiments, we emphasise that the passages in these
collections are pre-defined, \emph{atomic} retrieval units. Systems are
never required to consider whether to retrieve all or part of a document /
passage.

The idea of using LLMs to generate both relevance labels and accompanying
explanations for Q-D pairs was studied in \cite{exaranker}, but
with a different objective in mind. Retrieval datasets were augmented with
synthetic, LLM-generated explanations. A neural ranking model (ExaRanker)
trained on a small number of examples and their accompanying explanations
performed at par with models fine-tuned on many more examples, but without
explanations. In contrast, we consider this issue from a combination of the
LLMs-as-assessors and Explainability perspectives, and need LLMs to
generate extracts rather than synthesise explanations.

Extracting relevant passages may also be potentially applied to explaining
Retrieval-Augmented Generation (RAG) systems~\cite{patrick-lewis-rag,
  lee-etal-2019-latent}. Specifically, our approach can help ground the
LLM’s generated responses or quantify the degree of overlap between the
retrieved content and the ground-truth information. The work
by~\cite{rag_benchmark} introduces a benchmark in this direction,
highlighting the contextual cues present in the document as identified by
the LLM itself. Their evaluation measures the agreement between the LLM and
human annotators, rather than directly leveraging human-provided
annotations. We argue that incorporating human judgments when preparing
contextual tokens can help build a more reliable benchmark.

\vspace{-4mm}
\section{Our Approach}
\label{sec:approach}

\subsection{Background: the INEX Wikipedia collection}
\label{sec:inex}
The objective of the ad hoc track at INEX~\cite{inex_2009,inex_2010}, the
Initiative for the Evaluation of XML Retrieval, was \emph{focused
  retrieval}: given a collection of semi-structured (XML) documents, and a
query, to return a list of document \emph{excerpts} 
that precisely address the user's information need.
Ground truth was created by asking assessors to mark or \textbf{highlight}
``all, and only, relevant text in a pool of documents''~\cite{inex_2009}
retrieved by participating systems for each query. Systems were evaluated
on the basis of their ability to return all, and only, highlighted text.
The INEX 2009 and 2010 ad hoc test collections were constructed using a
Wikipedia dump from October 2008; user queries covered a variety of topics
of general interest. In order to ascertain whether LLMs-as-judges agree
with humans about which parts of a relevant document are actually useful,
we naturally turned to these collections.


\subsection{Prompt structure}
\label{sec:prompt}

\begin{figure}[b]
  \centering \includegraphics[scale=0.8]{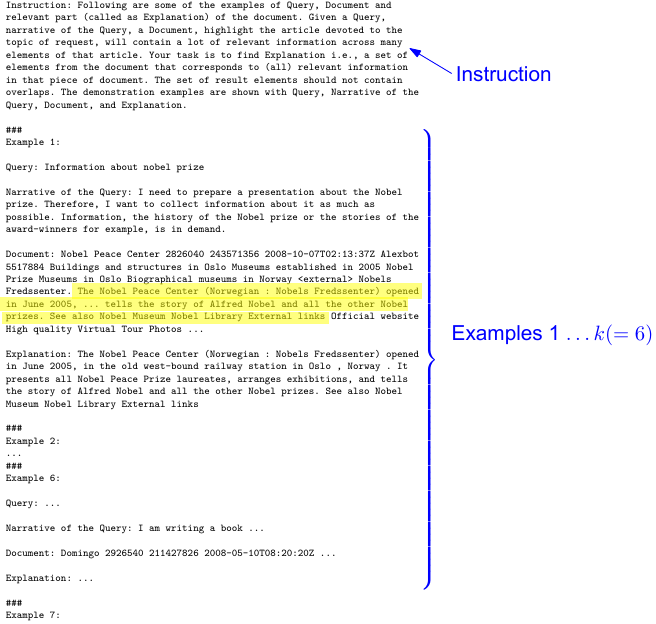}
  \caption{Structure of our prompt. For Example 7, which corresponds to the
    test instance, the query description, narrative and document are
    provided; the LLM is instructed to provide the relevant excerpts from
    the document. We highlight the rationale (relevant passage) within the
    document for the reader's convenience.}
  \label{fig:prompt}
\end{figure}

Thanks to earlier work in this
area~\cite{llm_judges_bhaskar_da,umbrela-ictir-25,lara}, a well-established
prompt structure has emerged for LLMs-as-judges in IR. Instead of spending
scarce computational and financial resources on additional prompt
engineering, we adopt the same general approach as used in recent work,
only modifying it to obtain highlighted passages or rationales from LLMs.
Following convention, our $k$-shot prompt starts with a description of the
task to be performed,
and then includes $k$ exemplars to illustrate what we want. In the
literature, this process is referred to as \textit{In-Context Learning
  (ICL)}~\cite{chowdhery2022palmscalinglanguagemodeling, few-shot}, and
does not involve updating model parameters; $k$ is determined by the prompt
size budget, and is set to 6 in this study.

Formally, let $\mathcal{E} = \{(x_i, y_i)\}_{i=1}^k$ denote the exemplars,
with each $(x_i, y_i)$ representing the $i^{th}$ input-output pair. In our
setup, each $x_i$ consists of ~(i)~the query description
($Q^{\text{desc}}$), ~(ii)~the query narrative ($Q^{\text{narr}}$), and
~(iii)~the document content ($D$); $y_i$ (equivalently $D^{\text{exp}}$) is
an excerpt corresponding to the highlighted portion(s) of $D$, that is
constructed by concatenating all passages within the document that are
marked as being relevant for the given query. The important components of a
prompt are shown in Figure~\ref{fig:prompt}.

The test instance, denoted $(x_{\text{test}}, y_{\text{test}})$, has the
same structure as an input-output pair. 
%
The LLM's prediction, $\hat{y}$, is given by
\[
\hat{y} = \text{LLM}[(x_1, y_1) \oplus \ldots \oplus (x_k, y_k) \oplus x_{\text{test}}],
\]
where $\oplus$ denotes concatenation.
These predictions are generated from the LLM using a greedy decoding strategy~\cite{ye-etal-2023-complementary}. We stop generating the sequence once we reach a
predefined maximum length. 
We then apply post-processing to the predicted output, $\hat{y}$, as
in~\cite{purohit2025sample}: regular-expression matching and filtering techniques
are employed to remove irrelevant segments of the generated output (e.g.,
prefixes such as ``\texttt{Explanation:}''). Finally, the ground-truth
highlighted explanation $y_{\text{test}}$ is compared with the
corresponding LLM-generated explanation $\hat{y}$ to compute recall and
precision. This is described in detail in Section~\ref{sec:main-results}. 

\subsection{Exemplar selection}
\label{sec:exemplar-selection}
Some earlier studies (e.g.,~\cite{lu-etal-2022-fantastically} on sentiment
classification tasks) have demonstrated that the selection and ordering of
exemplars in $\mathcal{E}$ may play a crucial role in determining the
accuracy of the predictions. Recently, \citet{craig-ecir-25-icl} have
considered the exemplar selection problem in the context of document-level
relevance prediction. However, their experiments were limited to a single
open-source model (Llama-3.8B-Instruct). They compared seven selection
strategies, and concluded that, for reliability, ``context examples should
be selected from the same query as the query to be judged and that many
context examples should be used.'' 

Our intention is to do a fine-grained assessment of LLMs-as-assessors in a
\textbf{few-shot} setting, i.e., we investigate whether, given only a few
query-highlighted-document pairs, LLMs can mimic humans when assessing
documents for a sizeable collection of queries. For this scenario,
providing examples for each query to be judged is not practical. For our
initial experiments, we randomly select Q-D pairs as exemplars.
Subsequently, we consider more structured approaches based on stratified
sampling to do exemplar selection (see Section~\ref{exemplar-strategy} for details).

\subsection{Evaluating LLM output}
\label{sec:eval}
Our task prompts an LLM to produce a single (not necessarily contiguous)
relevant excerpt from the document for a given query-document pair. 
To assess the quality of the output produced by an LLM (denoted $\hat{y}$
in Section~\ref{sec:prompt}), we need to measure the overlap between the
LLM-generated output and ground truth in terms of precision and recall,
which respectively indicate the relevance of the generated content, and the
completeness of the information captured by the LLM.


We may view the original document $D$ as a sequence of characters. The
ground truth, $D^{\text{exp}}$, is \emph{necessarily} a subsequence of $D$.
Because $\hat{y}$ represents content generated by an LLM, it is not
guaranteed to have this property. In principle, an LLM could paraphrase
document content (e.g., to increase succinctness or clarity, as judged by
the LLM itself), or even synthesize text that is semantically a subset of
the document content without corresponding to any easily identifiable
subsequence. We hope to avoid such scenarios by setting the `temperature'
parameter of recent LLMs to 0 during rationale generation, as this is expected
to minimise the `creativity' or `inventiveness' of the LLM. Further, since
all the exemplars provided to the LLM are subsequences, we may reasonably
expect that the output will also be in the same form. Section~\ref{verifying-llm-op}
provides more details about the extent to which these expectations are
fulfilled. 

To measure precision and recall, we first need to map each $\hat{y}$ to one
or more substrings of the corresponding document, by identifying the
longest common subsequence (LCS) 
between the document and $\hat{y}$. 
Once the LLM's output for a Q-D pair is mapped to subsequences
of the document, it is straightforward to compute the overlap between
ground truth annotations and the LLM's output, and in turn, recall,
precision and F$_1$ scores.


\subsection{Predicting relevance at the document level}
\label{sec:doc-level-rel}
To contextualise our findings with respect to related work, we also look at
document-level relevance prediction. In principle, this task could be
subsumed by the rationale-generation task. A document for which no
rationale is generated would then be labelled non-relevant. However, we
feel that this formulation of the task makes it unnecessarily difficult.
Any prompt designed for this purpose may be at least somewhat suggestive of
a trick-question. Thus, in addition to generating rationales (described
above in Section~\ref{sec:prompt}), we adopt the general approach that is
commonly used~\cite{llm_judges_bhaskar_da,umbrela-ictir-25,lara, kojima},
to get LLMs to predict document-level relevance. These experiments provide
the necessary background against which the results of rationale generation
should be interpreted.

The specific difference with the method for rationale generation outlined
in Section~\pageref{sec:prompt} is that we use a zero-shot,
instruction-only prompt. The prompt includes only a query, its detailed
description and narrative, and a document, and asks the LLM to classify the
document as relevant/non-relevant. For open-source models, we obtain the
LLM-predicted class probabilities, whereas for closed-source models, we
receive a simple \textsc{yes/no} response indicating whether the document
is relevant to a particular query.






\section{Experimental setup}
\label{sec:setup}

\subsection{Test collections}
\label{sec:test-collections}
As mentioned in Section~\ref{sec:inex}, we use the INEX 2009 and 2010 ad
hoc track test collections for our experiments. Both collections use the
same Wikipedia-based document collection, consisting of 2,666,190 articles.
The original INEX collection can be downloaded
from here: \href{https://www.mpi-inf.mpg.de/departments/databases-and-information-systems/software/inex/}{[INEX Homepage]}.
To the best of our knowledge,\footnote{Based on personal communication with
  the organizers of INEX.} this is currently the only available source for
the INEX benchmark collection. Table~\ref{tab:data-set-stats} presents a
brief summary of some statistics about these datasets; additional details
can be found in~\cite{inex_2009, inex_2010}.

For our experiments, we discard all XML markup and semantic annotations
from the Wikipedia articles, and extract the plain textual content using a
standard XML parser (libxml2). A total of 115 and 107 queries
(\emph{topics}, in TREC terminology) were created for INEX 2009 and 2010,
respectively, but out of these, 68 queries were judged for INEX 2009, while
52 queries were judged for INEX 2010. We use these 68 + 52 queries for our
experiments. 

The document pools were constructed from 172 runs submitted to INEX 2009,
and 148 submissions for INEX 2010. A total of 4,858 and 5,471 relevant
articles, respectively, were found relevant for the 68 + 52 queries that
were judged for INEX 2009 and 2010. Out of the 4,858 relevant documents in
the INEX 2009 collection, 3,339 (just under 69\%) contain only a single
highlighted passage; the corresponding figure for INEX 2010 is 3,388
documents out of 5,471 (about 62\%).

\begin{table}[h]
    \centering
    \caption{Statistics of the INEX adhoc datasets used in our experiments.}
    \label{tab:data-set-stats}
    \begin{tabular}{l c c c}
    \toprule
    \textbf{Dataset} & \textbf{\# Topics} & \textbf{\# Rel.\ docs} & \textbf{\# Docs with single}\\ 
    & & &  \textbf{rel. chunk} \\
    \midrule
    INEX 2009 & 68 & 4,858 & 3,339 \\ 
    INEX 2010 & 52 & 5,471 & 3,388 \\ 
    \bottomrule
    \end{tabular}
\end{table}

\subsubsection*{Ground truth}
The INEX collection uses binary judgments.
The ground truth (`qrels') provided by INEX is based on the fact that,
given a text file (either with or without markup), any contiguous
highlighted passage can be unambiguously identified via its starting byte
offset and length. The qrel file contains one line for each judged
Q-D pair; an example is shown in Figure~\ref{fig:qrel}. Apart
from the query and document identifiers, the line contains the total amount
of highlighted text in bytes, and a list of one or more $\langle$starting
offset, length$\rangle$ pair(s) corresponding to the highlighted
passage(s).\footnote{A non-relevant document can be easily identified
  because the total amount of highlighted text in it is 0.}
\begin{figure}[h!]
  \centering
  2009001 Q0 1528075 49158 58542 126 126:28761 28893:20397
  \caption{A line from an INEX 2009 qrel file}
  \label{fig:qrel}
\end{figure}

\subsection{Evaluation metrics}
When presenting aggregate recall, precision and F1 figures in the next
section, we compute averages at both the macro and micro levels. The
micro-averages are calculated by directly averaging over \emph{all}
Q-D pairs (4,858 pairs for INEX 2009, and 5,471 pairs for INEX
2010), without grouping these pairs by query. For macro-level evaluation
(see Equations~1 and 2), we first calculate the average value of a measure
across all Q-D pairs for each query $Q$. The mean of these
values across all judged topics, i.e., of 68 values for INEX 2009, and 52
values for INEX 2010, is reported.
\begin{eqnarray}
\label{eq:macro}
\text{Precision}_{\text{macro}} = \frac{1}{|Q|} \sum_{q=1}^{|Q|} \text{Precision}_q \\
\text{Recall}_{\text{macro}} = \frac{1}{|Q|} \sum_{q=1}^{|Q|} \text{Recall}_q
\end{eqnarray}
($|Q|$ denotes the total number of judged topics).

\subsection{Models used}
For extracting relevant / explanatory segments from documents, we employed
Llama 3.1
8B-Instruct\footnote{\url{https://ai.meta.com/blog/meta-llama-3-1/}},
GPT-4.1-mini\footnote{\url{https://platform.openai.com/docs/models/gpt-4.1-mini}},
and
GPT-4o-mini\footnote{\url{https://platform.openai.com/docs/models/gpt-4o-mini}}.
As of January, 2026, OpenAI \url{https://platform.openai.com/docs/models},
describe GPT-4.1 as their smartest non-reasoning model, with GPT-4.1 mini
being a ``smaller, faster version of GPT-4.1'', and GPT-4o-mini having
reasoning capabilities on several key reasoning benchmark
tasks\footnote{\url{https://openai.com/index/gpt-4o-mini-advancing-cost-efficient-intelligence/}}.
While GPT-5.1 has also become available, it is recommended for more complex
tasks, e.g., coding, that may need agentic involvement. The mini models
thus represent a good choice for us. We believe our selection of models is
consistent with that in earlier studies, 
even in these fast moving times. It is particularly important to note that
our experiments require LLMs to generate orders of magnitude more tokens
than simple \textsc{yes/no} relevance judgments, and are therefore
significantly more expensive. This precludes, for now, the possibility of
including a wider variety of models across providers.

Following standard practice~\cite{lara}, for the document-level relevance
prediction task, the parameters of Llama-3.1-8B-Instruct were
\textit{top-}$k = 50$ and \textit{temperature} $= 3.0$. For explanation
extraction using this smaller model, the settings were \textit{temperature}
$= 1.0$, \textit{top-}$k = 50$, and \textit{top-}$p = 1.0$. For
GPT-4.1-mini and GPT-4o-mini, a much larger and potentially more creative
model, we set the \textit{temperature} to $0$ and \textit{top-}$p$ to $1$,
enabling more deterministic outputs.


\section{Experimental results}
\label{sec:results}
An explainable LLM-based relevance assessment (RA) system might employ the
following two-stage pipeline. First, a $Q$-$D$ pair is given to an LLM that
is asked to predict whether $D$ is relevant for $Q$. If the answer is yes,
the LLM (or possibly a different LLM) is asked to explain its prediction by
providing precise excerpts from $D$ that are specifically relevant for $Q$.
In this section, we present experimental results that indicate how
effective LLMs might be in completing these two stages.

\subsection{Document-level relevance prediction}
We first present results for the document relevance prediction task. For
these experiments, we use only a subset of the complete test data, since
rationale generation (Section~\ref{sec:main-results}) is really our primary
focus, and therefore the main target of our computational and financial
resources. We have LLMs predict relevance for all the known relevant
documents (these are later used in our rationale generation experiments).
To provide a counterpoint, we also need to have LLMs predict relevance for
non-relevant documents. One standard approach would be to consider all
documents that were included in the pool and judged to be non-relevant.
Because the number of such documents is too large (given our constraints),
we select a smaller set of non-relevant documents as follows. We start with
the INEX ad hoc track submissions ($172$ for 2009 and $148$ for 2010), and
apply a Borda-count based rank aggregation algorithm to obtain a single
ranked list. Non-relevant documents that appear in this merged list between
ranks [1–100], [400–500] and [700–800] are selected to form Non-rel
Group-1, Non-rel Group-2 and Non-rel Group-3 respectively. Naturally, the
number of non-relevant documents thus selected will vary across both
queries and groups.

\subsubsection{Prediction results on the relevant document set}
Table~\ref{tab:rel-predict} reports prediction accuracy on the relevant
document sets from the INEX 2009 and 2010 collections, using Llama-3.1-8B, GPT-4.1-mini,
and GPT-4o-mini.\footnote{For the Llama-3.1-8B model, a document is
  labelled relevant if the LLM-predicted probability for a query document pair
  (Section~\ref{sec:doc-level-rel}) is at least $0.5$. The GPT-4.1-mini and GPT-4o-mini models do not provide probability scores for their predictions.}
Interestingly, among the three models,
the most `advanced' (GPT-4o-mini) 
exhibits the weakest predictive performance. GPT-4.1-mini does better, but
it does not match the performance of the simplest Llama-3.1-8B model. 
This suggests that GPT-4.1-mini and GPT-4o-mini tend to be very strict
when determining relevance. Potential reasons for this behavior
are explored further in the next section, where we present a
fine-grained analysis of LLM-based judgments.

\begin{table}[h]
    \centering
    \caption{Accuracy of LLMs-as-Judges in document relevance prediction on the \textit{\ul{relevant set}} for the INEX 2009 and INEX 2010 datasets.}
    \label{tab:rel-predict}
    \begin{tabular}{l c c}
    \toprule
    \textbf{Models} & \textbf{INEX 2009} & \textbf{INEX 2010} \\
    \midrule
    Llama-3.1-8B &  0.9035        & 0.9375        \\    
    GPT-4.1-mini &  0.6099        & 0.7213         \\   
    GPT-4o-mini  &  0.5683	&  0.6353                \\
    
    \bottomrule
    \end{tabular}
\end{table}


\subsubsection{Prediction results on the non-relevant document set}
Table~\ref{tab:rel-predict-non-rel-sampled} presents the prediction
accuracy of the same three models
across three groups of non-relevant documents.\footnote{Note that the
  figures in Tables~\ref{tab:rel-predict} and
  \ref{tab:rel-predict-non-rel-sampled} correspond to \emph{single}
  datapoints for the entire dataset, rather than an average over multiple
  datapoints. Thus, standard tests of significance cannot be applied to
  these figures.}

The results are worse than random for Llama-3.1-8B model, which produces
an enormous number of false positives. In contrast, GPT-4.1-mini and
GPT-4o-mini exhibit substantially higher accuracy across all three groups
for both the INEX 2009 and INEX 2010 collections. These results indicate
that Llama-3.1-8B may struggle to correctly identify non-relevant
documents, particularly those that are ranked highly by traditional
retrieval systems. Even though these results are somewhat disappointing,
they are consistent with those reported in \cite{alofi}. We also observe,
as in Table~\ref{tab:rel-predict}, that increasingly capable models are
more conservative when predicting relevance. Thus, GPT-4o-mini more
frequently predicts documents as non-relevant compared to the other models
considered in this study. Finally, as we move down the ranking, performance
improves consistently across models, as expected.

\begin{table}[h]
    \centering
    \caption{Accuracy of document relevance prediction using Llama-3.1-8B
      on the sampled \textit{\ul{non-relevant}} articles from the INEX 2009
      and INEX 2010 datasets.}
    \label{tab:rel-predict-non-rel-sampled}
\resizebox{1\columnwidth}{!}{

    \begin{tabular}{l c c c}
    \toprule
     \multirow{2}{*}{\textbf{Models}} & \textbf{Group-1} & \textbf{Group-2} & \textbf{Group-3} \\
      & \textbf{ranks [1-100]} & \textbf{ranks [400-500]} & \textbf{ranks [700-800]} \\

    \midrule
                    & & \textbf{INEX 2009} & \\\midrule
    Llama-3.1-8B  &  0.3005  &  0.4191  &  0.4776           \\
    GPT-4.1-mini &  0.7755  & 0.8565   & 0.8825   \\
    GPT-4o-mini  & 0.8060    &  0.8672  & 0.8966   \\\midrule
                    & & \textbf{INEX 2010} & \\\midrule
    Llama-3.1-8B &  0.2955  &  0.4021  &  0.4586           \\
    GPT-4.1-mini  &  0.7962  & 0.8594   & 0.8914   \\
    GPT-4o-mini  & 0.8327    &  0.8836  & 0.9047   \\


    \bottomrule
    \end{tabular}
    }
\end{table}




\subsubsection{Impact of lexical cues on false-positive on non-relevant set}
Next, we examine the false-positive predictions of the Llama-3.1-8B model
on the non-relevant document set in greater detail. We select Group-1
(non-relevant documents ranked between 1 and 100)
to examine the impact of lexical cues. We compute the BM25 score ($k_1 =
1.2$, $b = 0.75$) for each document (with respect to the combined title and
description of the query) as a measure of its lexical similarity with the
query. We then find the correlation between the Llama-3.1-8B–predicted
relevance probability scores for this set and the corresponding BM25 scores
The Spearman correlation ($\rho$) for the INEX 2009 and 2010 collections is
0.2588 and 0.2843, respectively, while Kendall’s $\tau$ is 0.1754 and
0.1962. At a high level, these results suggest that false-positive
predictions by LLMs are not strongly associated with lexical cues ---
specifically, the presence of query terms in documents, contrary to the
findings of Alaofi et al.~\cite{alofi}. Because the other LLMs do
not provide generation probabilities, we cannot subject them to a similar 
analysis.


\subsection{Main results: extracting relevant passages}
\label{sec:main-results}
Next, we turn to our main results, which focus on the fine-grained
evaluation of LLMs-as-Judges via rationale generation. As described in
Section~\ref{sec:approach}, we employ in-context learning (ICL) to identify
and highlight the relevant portions of each document in a Q-D
pair. Document lengths vary substantially in the Wikipedia corpus: for
instance, the smallest relevant document in INEX 2009 is 204 bytes, while
the largest is 159,845 bytes. To account for this
  variation, we construct three distinct sets of exemplars based on the
length of the document. In the first exemplar set (Exemplar$_1$), the
documents are very short; in the second set (Exemplar$_2$), longer
documents are randomly sampled for inclusion in the prompt; and in the
third set (Exemplar$_3$), the documents are of intermediate length. We fix
the exemplar budget to $k = 6$ for all three exemplar sets. Given the excessively conservative behaviour of
  GPT-4o-mini in document-level prediction, we begin our study with
  GPT-4.1-mini. The precision (P), recall (R), and F$_1$ scores at both
the macro and micro levels are reported in Table~\ref{tab:llm_performance} for the
Llama-3.1-8B and GPT-4.1-mini models.

\begin{table*}[h]
\centering
\caption{Performance of three exemplar configurations (Exemplar$_1$, Exemplar$_2$, and Exemplar$_3$) on the INEX 2009 and INEX 2010 datasets for Llama-3.1-8B and GPT-4.1-mini, respectively. We report both macro- and micro-level evaluations using Precision (P), Recall (R), and F$_1$ scores on these two collections. For each evaluation, the best-performing LLM (with the corresponding exemplar) is highlighted in bold. Underlined figures are discussed later. }
\label{tab:llm_performance}
\resizebox{\textwidth}{!}{
\begin{tabular}{l ccc  ccc  ccc  ccc}
\toprule
\multirow{2}{*}{\textbf{Model}} & \multicolumn{6}{c}{\textbf{INEX 2009}} & \multicolumn{6}{c}{\textbf{INEX 2010}} \\
\cmidrule(lr){2-7} \cmidrule(lr){8-13}
 & \multicolumn{3}{c}{\textbf{Macro}} & \multicolumn{3}{c}{\textbf{Micro}} & \multicolumn{3}{c}{\textbf{Macro}} & \multicolumn{3}{c}{\textbf{Micro}} \\
\cmidrule(lr){2-4} \cmidrule(lr){5-7} \cmidrule(lr){8-10} \cmidrule(lr){11-13}
 & P & R & F$_1$ & P & R & F$_1$ & P & R & F$_1$ & P & R & F$_1$ \\
\midrule
Exemplar$_1$ (Llama-3.1-8B) & 0.4501 & 0.6793 & 0.4997 & 0.5634 & 0.6681 & 0.6113 & 0.3547 & 0.6658 & 0.5136 & 0.5172 & 0.6480 & 0.5753  \\
Exemplar$_1$ (GPT-4.1-mini) & \textbf{0.6231} & 0.7146 & 0.6415 &  \ul{\textbf{0.6185}} & 0.7162 & 0.6638 & 0.5774 & 0.7425 & \textbf{0.6234} & 0.6117 & 0.7494 & \textbf{0.6736} \\

Exemplar$_2$ (Llama-3.1-8B) & 0.4622 & 0.7152 & 0.5176 & 0.5226 & 0.7055 & 0.6004 & 0.3684 & 0.6912 & 0.5312 & 0.4691	& 0.6962	& 0.5605  \\
Exemplar$_2$ (GPT-4.1-mini) & 0.5740 & \textbf{0.7988} & \textbf{0.6432} &  0.5784 & \textbf{0.8091} & \textbf{0.6746} & 0.5074 & \textbf{0.8263} & 0.6019 & 0.5282 & \ul{\textbf{0.8364}} & 0.6475 \\

Exemplar$_3$ (Llama-3.1-8B) & 0.3967 & 0.5945 & 0.4387 & 0.5052	& 0.6096  & 0.5525 &  0.3371 & 0.6079 & 	0.4472 &   0.4572	&  0.5962	&  0.5175  \\
Exemplar$_3$ (GPT-4.1-mini) & 0.6210 & 0.6644 &  0.6182 & 0.6107 & 0.6664 & 0.6373 & \textbf{0.5887} & 0.6753 & 0.6080 & \textbf{0.6142} & 0.7023 & 0.6553 \\

\bottomrule
\end{tabular}
}
\end{table*}

\subsubsection{\textbf{Verifying LLM generated output}}
\label{verifying-llm-op}
Before presenting the main results, we analyze the extent to which the
model extracts content verbatim. Portions of the generated explanations
that do not appear in the original document correspond to paraphrased
content, or possible ``hallucinations''. An examination of the longest
common subsequences between the generated explanations and the ground truth
output confirm that these possible paraphrases / hallucinations are not a
significant concern. For example, for GPT-4.1-mini on the 2009 collection,
the ``hallucination'' was less than 100 characters for
13,036/14,574\footnote{The total number of Q-D pairs for INEX 2009 is
  14,574 (3 exemplars $\times$ 4,858 relevant documents)} Q-D pairs, and
less than 10 characters for 10222 / 14,574 Q-D pairs.

\subsubsection{\textbf{Performance on the smaller LLM}}
We begin our analysis with the results obtained using the smaller model, Llama-3.1-8B. For this model, the maximum generation length is configured to match the length of each document. 
Overall, Exemplar$_2$ yields the highest precision and recall scores across
both the INEX 2009 and INEX 2010 collections, consistently outperforming
the other exemplar configurations across all macro- and micro-level
measures. 
Specifically, at the macro level, the precision, recall, and F$_1$ scores are $0.4622$, $0.7152$, and $0.5176$ for INEX 2009, and $0.3684$, $0.6912$, and $0.5312$ for INEX 2010. 
At the micro level, the corresponding values are $0.5226$, $0.7055$, and $0.6004$ for INEX 2009, and $0.4691$, $0.6962$, and $0.5605$ for INEX 2010. 

In general, the micro-level precision scores are higher than their macro-level counterparts. This difference arises because micro-level measures assign uniform weight across all Q-D pairs, whereas macro-level measures assign equal weight to each query. 
Consequently, queries with only a few relevant documents 
tend to reduce the overall macro-level performance. For the rest of our
analysis, therefore, we focus on micro-measures.

\begin{figure*}[t]
    \centering
    \begin{subfigure}[t]{0.40\textwidth}
        \centering
        \includegraphics[height=0.23\textheight, keepaspectratio]{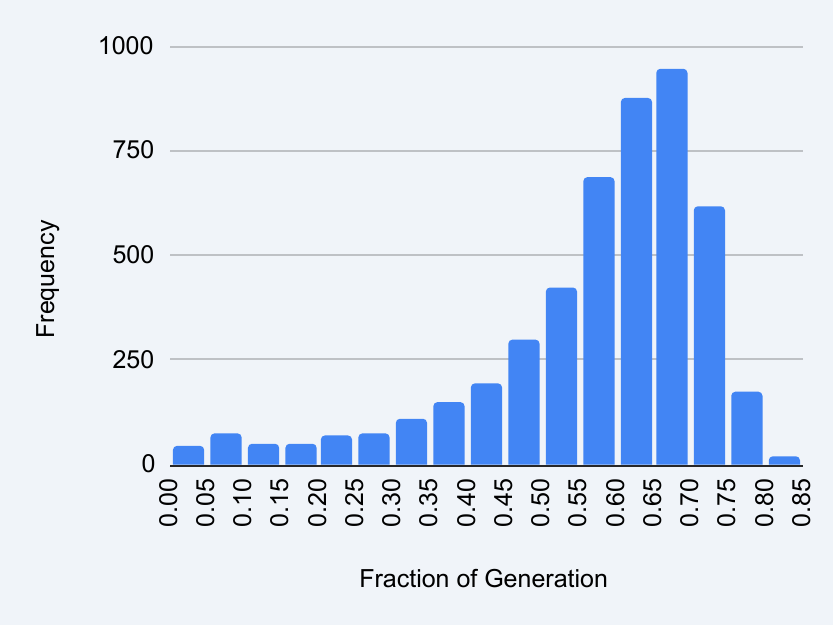}
        \caption{Histogram showing the distribution of the fraction of each document highlighted by Llama-3.1-8B for INEX 2009, using Exemplar$_2$.}
        \label{fig:hist_2009}
    \end{subfigure}
    \hfill
    \begin{subfigure}[t]{0.40\textwidth}
        \centering
        \includegraphics[height=0.23\textheight, keepaspectratio]{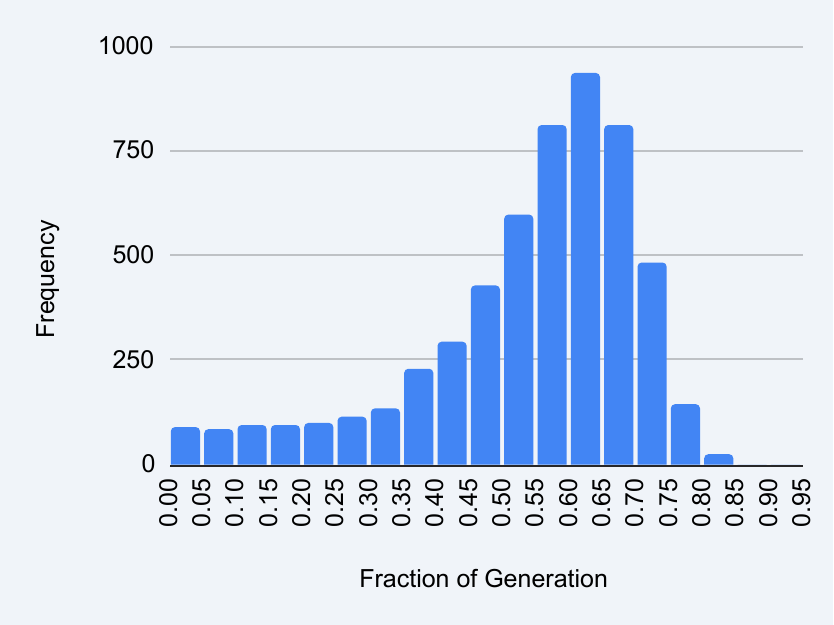}
        \caption{Histogram showing the distribution of the fraction of each document highlighted by Llama-3.1-8B for INEX 2010, using Exemplar$_2$.}
        \label{fig:hist_2010}
    \end{subfigure}
    \caption{Distribution of content lengths generated by Llama-3.1-8B relative to the corresponding document lengths in INEX 2009 and 2010, using Exemplar$_2$.}
    \label{fig:histograms}
\end{figure*}

It is noteworthy that the recall scores are generally higher than the precision values.
To further investigate this trend, we examine the ratio between the lengths
of the generated rationales and the corresponding full document lengths. 
Figure~\ref{fig:histograms} presents the histograms of these ratios for both collections using Exemplar$_2$. 
For INEX 2009, the model highlights more than half of the document in 3,922 out of 4,858 cases (approximately $80.7\%$), whereas for INEX 2010, this occurs in 3,808 out of 5,471 cases (approximately $69.6\%$). 
This observation indicates a tendency of the model to over-highlight document content, thereby inflating recall scores. 
In the limiting case where the model highlights the entire document, the recall would trivially reach $1.0$.

\subsubsection{\textbf{Performance on the larger model.}}
Next, we analyze the performance of the larger model (GPT-4.1-mini) for the three exemplar sets. 
Due to budget constraints, the maximum generation length for this larger model was limited to $8192$. 
This value was chosen based on the empirical distribution of ground-truth
highlighted chunk lengths, and ensures coverage of almost all observed
relevant chunks.
When transitioning from Llama-3.1-8B to GPT-4.1-mini, we observe a clear improvement in precision, which in turn leads to higher F$_1$ scores. 
Specifically, on INEX 2009, the highest macro-level precision increases from $0.4622$ to $0.6231$ (a relative gain of $34.8\%$), while the micro-level precision improves from $0.5226$ to $0.6185$ (an $18.3\%$ increase). 
A similar trend is observed for INEX 2010, indicating the expected performance gains with a larger and more parameter-rich model. 
Interestingly, while Exemplar$_2$ consistently outperformed the other configurations with Llama-3.1-8B, GPT-4.1-mini exhibits a different trend---Exemplar$_1$ and Exemplar$_3$ achieve comparable or even superior performance relative to Exemplar$_2$. 
This observation aligns with prior findings
by~\citet{lu-etal-2022-fantastically}, who argued that prompts effective
for smaller models may not necessarily yield the same benefits when scaled
to larger models (even within the same family).
The differences between Llama-3.1-8b and GPT-4-mini were found to be significant at the 95\% level (using a t-test) for all 3 exemplar sets, and all metrics, with only one exception: recall (macro-level) for Exemplar$_1$. 


\subsubsection{\textbf{Needle in haystack observation}}
\begin{figure*}[t]
    \centering
    \includegraphics[width=0.9\textwidth]{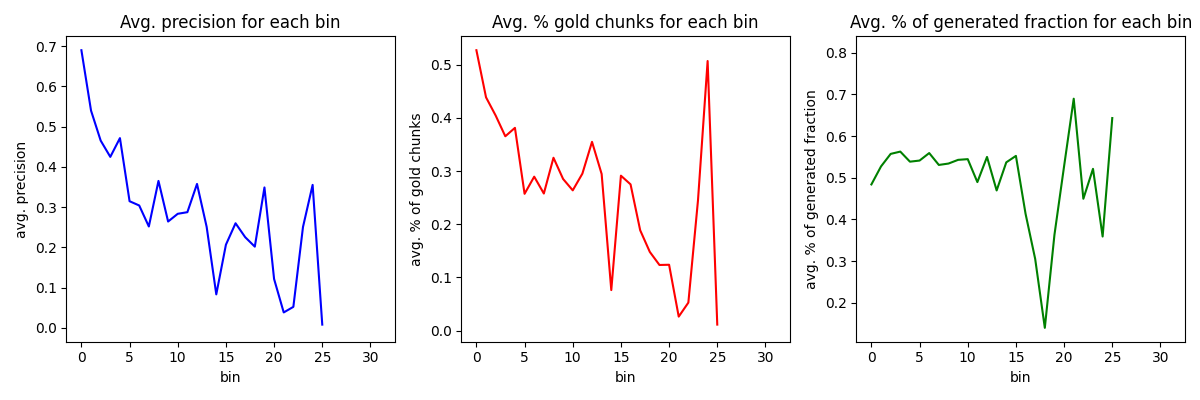}
    \caption{Distribution of document length vs precision, document length vs fraction of gold chunks, document length vs fraction of generated portions for 
    the INEX 2009 dataset using Exemplar$_2$ with Llama-3.1-8B model. The number of bins was determined using Rice’s Rule ($k = 2n^{1/3}$).}
    \label{fig:needle_haystack}
\end{figure*}

We further analyze the interaction between document length, the fraction of ground-truth relevant content, and the fraction of model-generated highlights with respect to precision. 
We sort all query-document pairs by document length and partition them into bins. 
For each bin, we compute and plot the precision score, the average fraction
of ground-truth relevant content, and the average ratio between generated
output and document length. 
Figure~\ref{fig:needle_haystack} presents these histograms for the INEX
2009 collection using Exemplar$_2$ (additional similar plots can be found
at \url{https://anonymous.4open.science/r/llm-as-judge}).
The number of bins was determined using \emph{Rice’s Rule}~\cite{rice-rule-1,rice-rule-2} as $k = 2n^{1/3}$, where $n$ is the number of observations.
As observed in Figure~\ref{fig:needle_haystack}, when only a small fraction of the document is relevant (e.g., bins $10$--$15$ and around $25$), the precision scores tend to be lower. 
This trend becomes particularly evident for longer documents, where the model struggles to identify the small relevant portions accurately.


\subsubsection{\textbf{Distribution of generated lengths vs.\ ground truth}}
\begin{figure}
    \centering
    \includegraphics[height=0.2\textheight,width=1\linewidth]{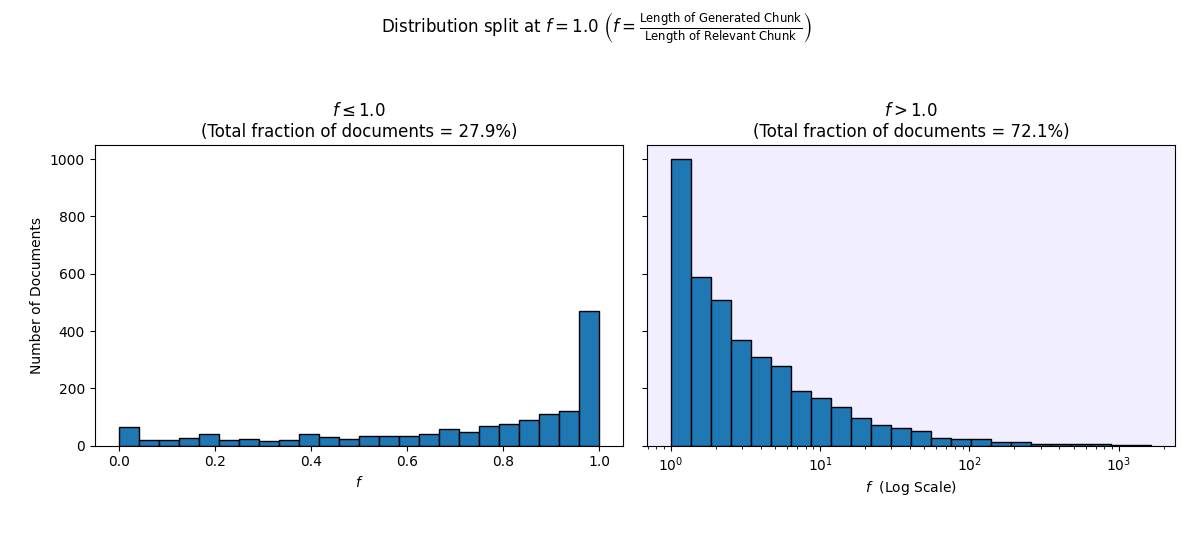}
    \caption{Distribution of the ratio between the generated chunk size and the relevant chunk size for Exemplar$_2$ using GPT-4.1-mini on the INEX 2010 dataset.}
    \label{fig:generated-vs-relevant-chunk}
\end{figure}

Figure~\ref{fig:generated-vs-relevant-chunk} shows the distribution of the
ratio between the generated chunk length and the relevant chunk length for
Exemplar$_2$ using GPT-4.1-mini on INEX 2010. A value greater than one
indicates that the model generates more content than the relevant chunk
highlighted by a human. This would typically lead to higher recall and
lower precision. The figure shows that the model generates longer chunks
than the relevant spans for approximately 72.1\% of the documents. Not
surprisingly, the average recall in this setting is 0.8364 (underlined in
Table~\ref{tab:llm_performance}). We observe a converse trend when average
precision is high, e.g., for Exemplar$_1$ on INEX 2009, this ratio is less
than one for approximately 51.1\% of the cases, indicating that the model
generates more concise outputs. These appear to better align with the
ground-truth relevant content, resulting in an average P of
0.6185.\footnote{This figure is not included because of space constraints,
  but may be found in our repository.}

\subsubsection{\textbf{Contiguous vs discontiguous chunks.}}
We divide the relevant set into two subsets: one containing documents in which the relevant content appears as a single, contiguous chunk, and the other containing documents having multiple, disjoint passages with relevant information. We observe that LLMs tend to struggle more when a document has discontiguous, relevant chunks. This behavior is consistent across both models. For instance, Llama-3.1-8B achieves a macro-level precision of $0.4691$ on the discontiguous set, while this increases to $0.5462$ on the contiguous set for INEX 2009 with Exemplar$_2$. Similarly, GPT-4.1-mini obtains macro-level precisions of $0.4951$ and $0.6163$ on the discontiguous and contiguous sets, respectively, using Exemplar$_2$. We observe similar trends across other exemplar sets, as well as on the 2010 collection.

\subsubsection{\textbf{Effect of larger model on hard data points}}

We analyze the Q-D pairs for which both LLMs (Llama-3.1-8B and
GPT-4.1-mini) perform poorly across all three exemplars; we refer to these
as hard data points. Our goal is to examine whether using a more capable
model can improve performance on this challenging subset. To identify such
instances, we select Q-D pairs for which the maximum precision achieved
across the six configurations (3 exemplars $\times$ 2 models) is at most
0.2. Additionally, we exclude pairs whose relevant chunk length is less
than 200 bytes.

This subset exhibits extremely low precision: for both collections, the average precision of Llama-3.1-8B and GPT-4.1-mini is near zero (Table~\ref{tab:llm_performance}, rows 1–2).
For these hard instances, we evaluate a reasoning-focused model, GPT-4o-mini, across all three exemplars. We observe consistent gains in precision: on INEX 2009, average precision increases to 0.1967, 0.1908, and 0.2520, and on INEX 2010 to 0.1490, 0.1534, and 0.2238. These results indicate that stronger reasoning capabilities can improve precision even on challenging cases. We next examine whether exemplar-selection strategies can further enhance performance on this subset.

\begin{table}[h]
\centering
\caption{Average precision measure of three exemplars (Exemplar$_1$, Exemplar$_2$, and Exemplar$_3$) denoted as (Ex$_1$, Ex$_2$, Ex$_3$) on the INEX 2009 and INEX 2010 datasets for Llama-3.1-8B and GPT-4.1-mini, and GPT-4o-mini on hard data points. \%improv. denotes the relative improvement of GPT-4o-mini over GPT-4.1-mini.}
\label{tab:llm_performance_hard_data}
\begin{tabular}{l ccc  ccc}
\toprule
\multirow{2}{*}{\textbf{Models}} & \multicolumn{3}{c}{\textbf{INEX 2009}} & \multicolumn{3}{c}{\textbf{INEX 2010}} \\
\cmidrule(lr){2-4} \cmidrule(lr){5-7}
 & Ex$_1$ & Ex$_2$ & Ex$_3$ & Ex$_1$ & Ex$_2$ & Ex$_3$ \\
\midrule
Llama-3.1-8B &  0.0607 & 0.0503 & 0.0620  & 0.0493 & 0.0434 & 0.0523  \\
GPT-4.1-mini &  0.0782 & 0.0674 & 0.0755  & 0.0728 & 0.0627 & 0.0750  \\
GPT-4o-mini  &  0.1967 & 0.1908 & 0.2520  & 0.1490 & 0.1534 & 0.2238   \\\midrule
\%improv. & 151.5 & 183.1 & 233.8 & 104.7 & 144.7 & 198.4 \\
\bottomrule
\end{tabular}
\end{table}

\subsubsection{\textbf{Exemplar selection strategy.}}
\label{exemplar-strategy}
We evaluate three exemplar-selection criteria: (i) average number of relevant documents per query (num-rel-docs), (ii) average relevant chunk size per query (rel-chunk), and (iii) average relevant chunk proportion per query (rel-chunk-prop). For all strategies, we restrict exemplars to single contiguous chunks, as previous experiments show that the models perform poorly with discontiguous chunks.
For each criterion, we partition queries into quartiles: below the median (Q$_1$), above the median (Q$_3$), and around the median (Q$_2$). We consider two sampling strategies: biased sampling, which selects Q-D pairs exclusively from Q$_1$ or Q$_3$, and balanced sampling, which samples uniformly across quartiles (Q$_{bal}$). In all settings, the exemplar budget is fixed to six.


Table~\ref{tab:llm_performance_hard_data_exemplars} reports results for these sampling strategies across the three criteria. We observe a statistically significant improvement (one-sided t-test, 95\% confidence level) when exemplars are selected from the Q$_1$ quartile for both collections. In particular, on INEX 2009, average precision increases from 0.2520 (Exemplar$_3$) to 0.3415 with rel-chunk-prop, and on INEX 2010, from 0.2238 to 0.2853.
\subsubsection{\textbf{Explaining the behavior of selection strategy.}}
The largest improvements are observed in the first quartile (Q$_1$) of rel-chunk-prop, where exemplars contain the smallest proportions of relevant content. These examples resemble a needle-in-a-haystack setting, in which limited relevance is embedded within substantial semantically related but non-relevant text, effectively teaching the model to isolate intent-matching evidence while ignoring surrounding noise. When relevant chunks are not extremely small in absolute size but occupy a small fraction of the document, the model learns to avoid over-attending to tangential or background information. In contrast, rel-chunk-size exhibits mixed behavior: very small relevant chunks may be too sparse to function as reliable relevance signals, while very large chunks may span broad document regions and encourage over-prediction of relevance. This explains the similar performance observed in the third quartile (Q$_3$) of both strategies. Balanced sampling shows limited benefit, as it does not prioritize low-proportion relevance cases and therefore fails to strongly emphasize the needle-in-a-haystack nature of the task.
Finally, num-rel-docs appears to be a weak signal for exemplar selection. The number of relevant documents per query is not directly observable by the model and primarily reflects corpus-level characteristics rather than document-level relevance structure. As a result, it provides limited guidance for identifying fine-grained relevance within individual documents.

\begin{table}[h]
\centering
\caption{
Average precision under different exemplar-selection strategies (num-rel-docs, rel-chunk-size, and rel-chunk-prop). Results are shown for Q$_1$, Q$_3$, and Q$_{bal}$ on the INEX 2009 and INEX 2010 datasets using GPT-4o-mini on hard data points. Baseline results with fixed exemplars (Ex$_1$, Ex$_2$, Ex$_3$) are included for comparison. Superscripts denote statistically significant improvements over the corresponding baseline (one-sided t-test, $p<0.05$).
}
\label{tab:llm_performance_hard_data_exemplars}
\begin{tabular}{c|ccc ccc}
\toprule
\multicolumn{7}{c}{\textbf{GPT-4o-mini with different exemplars}} \\
\cmidrule(lr){1-7}
 & Ex$_1$ & Ex$_2$ & Ex$_3$ & Q$_1$ & Q$_3$ & Q$_{bal}$ \\
\midrule

\multirow{5}{*}{\rotatebox{90}{\parbox[c]{1.2cm}{\centering \textbf{INEX 2009}}}}

& 0.1967 & 0.1908 & 0.2520 & NA & NA & NA \\

\cmidrule(lr){2-7}

& \multicolumn{3}{c}{num-rel-docs}
& 0.2836
& 0.2092
& 0.2049 \\

& \multicolumn{3}{c}{rel-chunk-size}
& 0.3448$^{Ex_3}$
& 0.1594
& 0.2407 \\

& \multicolumn{3}{c}{rel-chunk-prop}
& 0.3415$^{Ex_3}$
& 0.1623
& 0.2255 \\
\midrule


\multirow{5}{*}{\rotatebox{90}{\parbox[c]{1.2cm}{\centering \textbf{INEX 2010}}}}
& 0.1490 & 0.1534 & 0.2238 & NA & NA & NA \\
\cmidrule(lr){2-7}

& \multicolumn{3}{c}{num-rel-docs}
& 0.1861
& 0.2211
& 0.1939 \\

& \multicolumn{3}{c}{rel-chunk}
& 0.2131
& 0.1480
& 0.2229 \\

& \multicolumn{3}{c}{rel-chunk-prop}
& 0.2853$^{Ex_3}$
& 0.1262
& 0.2266 \\
\midrule


\bottomrule
\end{tabular}
\end{table}

\subsubsection{\textbf{Anecdotal analysis.}}
The INEX 2009 collection contains a single query that explicitly targets images of specific entities: ``Find images of sunflowers painted by Vincent van Gogh'' (query ID: 2009065). The narrative description of this query is shown below.
\begin{quote}
Being a Dutch post-impressionist artist, Vincent van Gogh produced many paintings that are now very popular, well-known, and valuable. His painting style has significantly influenced the development of modern art. Among his works, I particularly appreciate the various depictions of sunflowers for their vibrant colors, which express emotions typically associated with the life of sunflowers. I would like to find relevant sunflower paintings as images in documents so that I can enjoy their simplistic beauty. To be relevant, the sunflower must be presented as an image and painted by Vincent van Gogh. Sunflower images created by others are irrelevant.
\end{quote}

As the collection consists exclusively of textual documents, image-based content is not explicitly handled. Consequently, successful identification of such references would represent a noteworthy capability of the LLMs as Assessors paradigm.
For this query, GPT-4.1-mini achieves a maximum precision of 0.4972, compared to 0.2462 for Llama-3.1-8B. In contrast, Llama-3.1-8B attains substantially higher recall (0.9039) than GPT-4.1-mini, whose highest recall is 0.6881.
\subsubsection{\textbf{Failure analysis.}} 
Recall that each test case is included in the prompt as Example 7 with a blank Explanation field. Each test case is included in the prompt as Example 7 with a blank Explanation field. Llama-3.1-8B occasionally generates an extraneous Example 8, which is removed during post-processing but may be interpreted as hallucination. This occurs for Exemplar$_2$ in 13.6\% of cases on INEX 2009 and 16\% on INEX 2010, with similar or higher rates for other exemplars. In contrast, GPT-4.1-mini exhibits no such behavior on either collection.
A finer-grained analysis of GPT-4.1-mini shows that, for a small subset of Q-D pairs, the model refuses to generate explanation spans from the document, instead producing a recurring response such as ``No relevant information \ldots''. For INEX 2009, this behavior occurs in 19, 173, and 340 cases for the three exemplars, respectively; for INEX 2010, the corresponding counts are 21, 153, and 237. Because the model still produces an output, our evaluation framework does not penalize this behavior. Understanding why the model confidently predicts the absence of relevant information for these Q-D pairs remains an open question.

\section{Conclusions and future work}
\label{sec:cfw}

In this study, we present a fine-grained evaluation of LLMs as assessors. While prior work has largely examined document-level relevance prediction, we additionally require LLMs to generate rationales by identifying passages that justify their decisions. At a high level, our findings echo prior observations~\cite{clark-dietz-arxiv-2025,Soboroff_2025,sakai-cikm-25-short}. GPT-4.1-mini consistently outperforms Llama-3.1-8B; however, both models tend to over-generate content, resulting in high recall. In contrast, GPT-4o-mini exhibits more conservative document-level predictions and performs better on hard instances explanation generations. We further show that principled exemplar-selection strategies can substantially improve the precision of generated explanations.
All reported results should be interpreted in the context of inter-annotator agreement; however, only limited agreement information is available for the INEX collections 
Due to budget limitations, we evaluate exemplar-selection strategies on a subset of INEX 2009 and 2010 rather than the full collections. Finally, our preliminary analysis indicates that detailed step-by-step prompting encourages paraphrasing rather than faithful extraction, an issue we plan to investigate more systematically in future work.
\begin{acks}
  
\end{acks}

\bibliographystyle{ACM-Reference-Format}
\bibliography{references}

\end{document}